\documentclass[aps,twocolumn,preprintnumbers,showpacs,showkeys,nofootinbib%
]{revtex4}
\usepackage{epsfig}
\usepackage{amssymb,amsmath,amsfonts,amsthm,graphicx,psfrag}
\graphicspath{{./Figures/}}                 
\newcommand{\noi}{\noindent}
\newcommand{\beq}{\begin{equation}}
\newcommand{\eeq}{\end{equation}}
\newcommand{\bea}{\begin{eqnarray}}
\newcommand{\eea}{\end{eqnarray}}
\newcommand{\Fig}[1]{Fig.~\ref{#1}}
\newcommand{\Tab}[1]{Table~\ref{#1}}
\newcommand{\Sec}[1]{Section~\ref{#1}}
\newcommand{\Eq}[1]{Eq.~(\ref{#1})}
\newcommand{\caa}{{\cal A}}

\newcommand{\tr}{\operatorname{Tr}}

\newcommand{\bc}{{\it bc~}}
\newcommand{\fc}{{\it fc~}}

\newcommand{\aleq}{\mbox{}_{\textstyle \sim}^{\textstyle < }}
\newcommand{\ageq}{\mbox{}_{\textstyle \sim}^{\textstyle > }}

\begin{document}
\preprint{HU-EP-09/63, ITEP-LAT/2009-18}

\title{$SU(2)$ lattice gluon propagator: continuum limit, 
finite-volume effects and infrared mass scale $m_\mathrm{IR}$}

\author{V.~G.~Bornyakov}
\affiliation{Institute for High Energy Physics, 142281, Protvino, Russia \\
and Institute of Theoretical and Experimental Physics, 117259 Moscow, Russia}

\author{V.~K.~Mitrjushkin}
\affiliation{Joint Institute for Nuclear Research, 141980 Dubna, Russia \\
and Institute of Theoretical and Experimental Physics, 117259 Moscow, Russia}

\author{M.~M\"uller--Preussker}
\affiliation{Humboldt-Universit\"at zu Berlin, Institut f\"ur Physik, 
Newton-Str. 15, 12489 Berlin, Germany}

\date{December 22, 2009}

\begin{abstract}
We study the scaling behavior and finite (physical) volume effects
as well as the Gribov copy dependence of the $SU(2)$ Landau gauge gluon
propagator on the lattice. Our physical lattice sizes range from
$(3.0~\mathrm{fm})^4$ to $(7.3~\mathrm{fm})^4$.  Considering
lattices with decreasing lattice spacing but fixed physical volume 
we confirm (non-perturbative) multiplicative renormalizability 
and the approach to the {\it continuum limit} for the renormalized
gluon propagator $D_{ren}(p)$ at momenta $|p|\, \ageq \, 0.6$ GeV . The
finite-volume effects and Gribov copy influence turn out small in this 
region. On the contrary, in the deeper infrared we found the Gribov copy
influence strong and finite-volume effects, which still require
special attention. The gluon propagator does not seem to be consistent with a
simple pole-like behavior $\sim (p^2+m_g^2)^{-1}$ for momenta 
$|p|\, \aleq \, 0.6$ GeV. Instead, a Gaussian-type fit works very well in this
region.  From its width - for a physical volume $(5.0~\mathrm{fm})^4$ - 
we estimate a corresponding infrared (mass) scale to be 
$m_\mathrm{IR} \sim 0.7~$ GeV.
\end{abstract}



\maketitle

\section{Introduction}
\label{sec:introduction}

The Landau (or Lorenz) gauge gluon and ghost propagators in pure 
Yang-Mills theories or in full QCD have attracted much interest 
for many years.

One reason for this interest is that the behavior of these propagators
in the infrared (IR) region has been related to gluon and quark 
confinement~\cite{Gribov:1977wm,Zwanziger:1991gz,Kugo:1979gm}. 
In particular, Zwanziger has argued that the gluon propagator should 
vanish in the IR limit~\cite{Zwanziger:1991gz,Zwanziger:2001kw, 
Zwanziger:2009je}, while the ghost dressing function should become singular.

Another reason for this interest is connected with the importance of
the momentum dependence of both propagators, especially in the (deep)
IR-region, for the phenomenological analysis of experimental
data. Many years ago Parisi and Petronzio \cite{Parisi:1980jy}
have pointed out that non-zero effective gluon mass  (or dynamically
generated gluon mass) $m_g$ is important to resolve some discrepencies
in low-energy tests of QCD, as, e.g., ratios of widths of $J/\psi$.
Since then a number of papers has been dedicated to phenomenological
studies of such processes as $J/\psi \to \gamma X$, $\gamma \to \pi^0$
transition, non-leptonic $B$ meson decays, etc., where a non-zero
value of the effective gluon mass $m_g$  plays a crucial role  (for
an incomplete list of references see, e.g., the recent papers 
\cite{Field:2001iu,Natale:2009uz}). 

Let us note that in order to obtain a reliable value of $m_g$ one 
needs to know the continuum gluon propagator $D(p)$ in the {\it deep 
infrared region}. The definition of the mass $m_g$ is based on the 
hypothesis of a pole-like behavior, i.e. $\sim 1/(p^2+m_g^2)$, of 
the gluon propagator at small momenta. In this case the effective 
gluon mass defines the {\it infrared mass scale} $m_\mathrm{IR}$.

Gauge-variant QCD Green functions may serve also as input to bound
state equations as Bethe-Salpeter or Faddeev equations for hadron 
phenomenology \cite{Alkofer:2000wg,Alkofer:2006jf,Eichmann:2008ef}. 
Moreover, at large momenta they should allow a determination of 
phenomenogically relevant parameters such as
$~\Lambda_{\overline{MS}}~$ or condensates $~\langle A^2 \rangle, 
\langle \overline{\psi} \psi \rangle~$ by fitting lattice data to 
continuum expressions obtained from operator product expansion 
and perturbation theory~\cite{Chetyrkin:2004mf,Chetyrkin:2009kh}
in a certain MOM scheme~\cite{Sternbeck:2007br,vonSmekal:2009ae}. 

\hspace{2mm}

The search for intertwined asymptotic gluon and ghost
propagator solutions of Dyson-Schwinger (DS) and functional
renormalization group (FRG) equations showed the existence
of infrared solutions exhibiting a power-like 
{\it scaling}
behavior~\cite{vonSmekal:1997is,vonSmekal:1997vx,Alkofer:2000wg,
Alkofer:2002aa,Fischer:2002hna,Fischer:2002eq,Alkofer:2004it,Fischer:2006vf}.
However, as has been pointed out in~\cite{Boucaud:2007va,Boucaud:2008ji,
Boucaud:2008ky,Aguilar:2008xm}, there are also regular 
so-called {\it decoupling}
solutions providing an IR-finite limit of both the gluon propagator and
the ghost dressing function.  In \cite{Fischer:2008uz} it has been argued,
that it seems to be a question of IR boundary conditions posed on the
ghost dressing function, what kind of solution one has to select.

\hspace{2mm}

On the lattice, over the last decade extensive studies of
the Landau gauge gluon and ghost propagators have been carried
out~\cite{Suman:1995zg,Cucchieri:1997dx,Leinweber:1998uu,Becirevic:1999uc,
Bonnet:2000kw,Bonnet:2001uh,Bloch:2002we,Bloch:2003sk,Bakeev:2003rr,
Sternbeck:2005tk,Bowman:2007du}.  In the meantime lattice computations
have reached lattice volumes even with a linear extension $O(10
\mathrm{fm})$ in order to discriminate between the above mentioned
IR solutions. In this way lattice QCD has been found to support the
{\it decoupling} solution~\cite{Cucchieri:2007md,Cucchieri:2007rg,
Cucchieri:2008fc,Bornyakov:2008yx,Bogolubsky:2009dc}.  This is true as
long as one relies on finite-box periodic boundary conditions and on a
gauge condition requiring the Landau gauge functional to take extrema as
close as possible to the global extremum. 

The Gribov copy influence still remains a serious problem in the lattice
calculations, at least, in the deep IR-region. We believe that the
correct gauge condition should require the Landau gauge functional to take
extrema as close as possible to the {\it global} extremum \footnote{For
recent alternative attempts see~\cite{vonSmekal:2008en,Maas:2009se,
Dudal:2009xh,Dudal:2009bf}.}.
Indeed,

\begin{itemize}

\item[(i)] a consistent non-perturbative gauge fixing procedure proposed
by Parrinello-Jona-Lasinio and Zwanziger (PJLZ-approach)
\cite{Parrinello:1990pm, Zwanziger:1990tn} presumes that the choice of
a unique representative of the gauge orbit should be through the {\it
global} extremum of the chosen gauge fixing functional;

\item[(ii)] in the case of 
lattice compact $U(1)$ gauge theory in the weak coupling 
(Coulomb) phase some of the gauge copies produce a photon propagator
with a decay behavior inconsistent with the expected zero mass
behavior \cite{Nakamura:1991ww,Bornyakov:1993yy,Mitrjushkin:1996fw}.
The choice of the global extremum permits to obtain the physical
- massless - photon propagator.

\end{itemize}

For all practical purposes the system of DS and/or FRG 
equations has to be truncated. The details of truncation influence 
the behavior of the Green functions especially in the non-perturbative 
momentum range around $~1 \mathrm{GeV}$, where the Landau gauge 
gluon dressing function exhibits a pronounced maximum. 
Therefore, reliable results from first principles
to compare with are highly welcome. On the lattice, for finite volumes 
such results can be obtained and directly compared with finite volume 
DS and FRG results~\cite{Fischer:2005ui,Fischer:2007pf}. 
To our knowledge, on the lattice a systematic continuum limit 
determination of the Landau gluon and ghost propagators for various 
fixed physical volumes in the momentum range 
$~0.3~\mathrm{GeV}~\aleq~p~\aleq~10~\mathrm{GeV}~$ is still missing. Such
an evaluation has to make sure that Gribov copy effects, lattice artifacts
and multiplicative renormalizability are sufficiently under control.

\vspace{2mm}

Here we present such a study for $SU(2)$ pure gauge theory as
a continuation of our investigation in~\cite{Bornyakov:2008yx}.
For gauge fixing we rely on the Landau gauge with a gauge condition
requiring the gauge fixing functional to take extrema as close
as possible to the global extremum. For this aim we employ the
simulated annealing algorithm~\cite{Bali:1994jg,Bogolubsky:2007pq,
Bornyakov:2008yx} in combination with non-periodic $~Z(2)~$ gauge
transformations (``$Z(2)$-flips'')~\cite{Bogolubsky:2007bw}. With
respect to the latter we search within all $~2^4=16~$ global $~Z(2)~$
Polyakov loop sectors.  We concentrate on the gluon propagator and will
present data for three different physical volumes up to $~\sim (7.3
\mathrm{fm})^4$. As done in a preliminary manner in \cite{Bloch:2003sk} 
we check for multiplicative renormalizability and provide results for 
the renormalized propagator and dressing function, respectively, 
which can be considered already to be continuum ones.  
In \Sec{sec:definitions} we introduce the observables to be computed.  
In \Sec{sec:details} some details of the simulation are given, 
whereas in \Sec{sec:gribov} we discuss the effect of improved gauge 
fixing.  In \Sec{sec:results} we present our numerical results. 
\Sec{sec:infrared} is dedicated to the discussion of the deep
infrared region and to the definition of an alternative infrared 
(mass) scale $m_\mathrm{IR}$.  Conclusions will be drawn in 
\Sec{sec:conclusions}.

\section{The gluon propagator: definitions}
\label{sec:definitions}

In order to generate Monte Carlo ensembles of non-gauge-fixed
$SU(2)$ gauge field configurations we use the standard plaquette 
Wilson action  
\bea
S & = & \beta \sum_x\sum_{\mu >\nu}
\left[ 1 -\frac{1}{2}~\tr \Bigl(U_{x\mu}U_{x+\mu;\nu}
U_{x+\nu;\mu}^{\dagger}U_{x\nu}^{\dagger} \Bigr)\right]\,, \nonumber \\
& & \beta = 4/g_0^2 \,.
\label{eq:action}
\eea
$g_0$ denotes the bare coupling constant,  
$U_{x\mu} \in SU(2)$ are the link variables. The latter 
transform under local gauge transformations $g_x$ as follows 
\beq
U_{x\mu} \stackrel{g}{\mapsto} U_{x\mu}^{g}
= g_x^{\dagger} U_{x\mu} g_{x+\mu} \,,
\qquad g_x \in SU(2) \,.
\label{eq:gaugetrafo}
\eeq
The standard definition~\cite{Mandula:1987rh} for the dimensionless
lattice gauge vector potential $\caa_{x+\hat{\mu}/2,\mu}$ is
\beq
\caa_{x+\hat{\mu}/2,\mu} = \frac{1}{2i}~\Bigl( U_{x\mu}-U_{x\mu}^{\dagger}\Bigr)
\equiv A_{x+\hat{\mu}/2,\mu}^a \frac{\sigma_a}{2} \,.
\label{eq:a_field}
\eeq
This definition, which is not unique, can influence the propagator 
results in the IR region, where the continuum limit is hard to control.

In lattice gauge theory the usual choice of the Landau gauge condition
is~\cite{Mandula:1987rh}
\beq
(\partial \caa)_{x} = \sum_{\mu=1}^4 \left( \caa_{x+\hat{\mu}/2;\mu}
  - \caa_{x-\hat{\mu}/2;\mu} \right)  = 0 \,,
\label{eq:diff_gaugecondition}
\eeq
which is equivalent to finding a local extremum of the gauge functional
\beq
F_U(g) = ~\frac{1}{4V}\sum_{x\mu}~\frac{1}{2}~\tr~U^{g}_{x\mu} \,,
\label{eq:gaugefunctional}
\eeq
with respect to gauge transformations $~g_x~$.  $V=L^4~$ denotes the 
lattice volume. The manifold consisting of Gribov copies providing local 
maxima of the functional (\ref{eq:gaugefunctional}) and a semi-positive 
Faddeev-Popov operator is called the {\it Gribov region} $~\Omega$, while
that of the global maxima is called the {\it fundamental modular region} (FMR)
$~\Lambda \subset \Omega$.  Our gauge fixing procedure is aimed to approach
$~\Lambda$.

The (unrenormalized) gluon propagator $~D~$ and its dressing function $~Z~$
are then defined (for $p \neq 0$)  by
\bea
D_{\mu\nu}^{ab}(p)
&=& \frac{a^2}{g_0^2} 
    \langle \widetilde{A}_{\mu}^a(k) \widetilde{A}_{\nu}^b(-k) \rangle  
    \nonumber \\
&=& \left( \delta_{\mu\nu} - \frac{p_{\mu}~p_{\nu}}{p^2} \right)
    \delta^{ab} D(p)\,, 
\label{eq:gluonpropagator}
\\ 
Z(p) &=& D(p)~p^2 \,,
\label{eq:gluondressing}
\eea
where  $\widetilde{A}(k)$  represents the Fourier transform of
the gauge potentials defined by \Eq{eq:a_field} after having fixed
the gauge. $a$ denotes the lattice spacing.  
The physical momenta $p$ are given by 
$p_{\mu}=(2/a) \sin{(\pi k_{\mu}/L)}, ~~k_{\mu} \in (-L/2,L/2]$.
For $p \ne 0$, one determines $D(p)$ according to \Eq{eq:gluonpropagator}
\beq
D(p) = \frac{1}{9} \sum_{a=1}^3 \sum_{\mu=1}^4 D^{aa}_{\mu\mu}(p) \,,
\eeq
whereas the ``zero momentum propagator'' $D(p=0)$ is defined as
\beq
D(0) = \frac{1}{12} \sum_{a=1}^3 \sum_{\mu=1}^4 D^{aa}_{\mu\mu}(p=0) \,.
\eeq

\section{Details of the simulation}
\label{sec:details}

We have performed Monte Carlo (MC) simulations at various $\beta$-values
between $\beta=2.2$ and $\beta=2.55$ for various lattice sizes $L$.
Consecutive configurations (considered to be statistically independent)
were separated by 100 sweeps, each sweep consisting of one local heatbath
update followed by $L/2$ microcanonical updates. In \Tab{tab:data_sets}
we provide the full information about the field ensembles used in this
investigation.
%
\begin{table}[h]
\begin{center}
\vspace*{0.2cm}
\begin{tabular}{|c|c|c|c|c|c|c|} \hline
$\beta$ & $a^{-1}$ [GeV] & $a$ [fm] & $~L~$ & $aL$ [fm] & 
$~N_{meas}~$ & $N_{copy}$ \\ \hline\hline

 2.20  & 0.938 & 0.210 &  14  & 2.94  & 400  &  48     \\ 
 2.30  & 1.192 & 0.165 &  18  & 2.97  & 200  &  48     \\
 2.40  & 1.654 & 0.119 &  26  & 3.09  & 200  &  48     \\ 
 2.50  & 2.310 & 0.085 &  36  & 3.06  & 400  &  80     \\ 
 2.55  & 2.767 & 0.071 &  42  & 2.98  & 200  &  80     \\ \hline\hline
                                         
 2.20  & 0.938 & 0.210 &  24  & 5.04  & 400  &  48     \\ 
 2.30  & 1.192 & 0.165 &  30  & 4.95  & 400  &  48     \\ 
 2.40  & 1.654 & 0.119 &  42  & 5.00  & 200  &  80     \\ \hline\hline

 2.30  & 1.192 & 0.165 &  44  & 7.26  & 200  &  80     \\ \hline\hline

\end{tabular}
\end{center}
\caption{Values of $\beta$, lattice sizes, number of measurements
and number of gauge copies used throughout this paper. For the 
values of the lattice spacing see \cite{Bloch:2002we} 
($1$ GeV${}^{-1} \simeq 0.197$ fm).
} 
\label{tab:data_sets}
\end{table}

For gauge fixing we employ the $Z(2)$ flip operation as discussed in
\cite{Bogolubsky:2007bw}. For completeness we repeat the main 
information. The method consists in flipping all link variables
$U_{x\mu}$ attached and orthogonal to a $3d$ plane by multiplying them
with $-1 \in Z(2)$. 
Such global flips are equivalent to non-periodic gauge
transformations and represent an exact symmetry of the pure gauge action
considered here. The Polyakov loops in the direction of the chosen links
and averaged over the $3d$ plane obviously change their sign. Therefore,
the flip operations combine 
the $2^4$ distinct gauge orbits (or Polyakov loop sectors) of strictly 
periodic gauge transformations into one larger gauge orbit.

The second ingredient is the simulated annealing (SA) method, 
which has been found computationally more efficient than the only 
use of standard overrelaxation (OR) \cite{Schemel:2006da,Bogolubsky:2007pq,
Bogolubsky:2007bw}.
The SA algorithm generates gauge transformations
$~g(x)~$ by MC iterations with a statistical weight proportional to
$~\exp{(4V~F_U[g]/T)}~$. The ``temperature'' $~T~$ is an auxiliary
parameter which is gradually decreased in order to maximize the
gauge functional $~F_U[g]~$.  In the beginning, $~T~$ has to be chosen
sufficiently large in order to allow traversing the configuration space
of $~g(x)~$ fields in large steps. 
As in Ref. \cite{Bogolubsky:2007bw} we have chosen $~T_{\rm init}=1.5.$
After each quasi-equilibrium
sweep, including both heatbath and microcanonical updates, $~T~$ has
been decreased with equal step size. The final SA temperature has been fixed
such that during the consecutively applied OR algorithm the violation of 
the transversality condition
\beq
\max_{x\mbox{,}\, a} \, \Big|
\sum_{\mu=1}^4 \left( A_{x+\hat{\mu}/2;\mu}^a - A_{x-\hat{\mu}/2;\mu}^a \right)
\Big| \, < \, \epsilon_{lor}
\label{eq:gaugefixstop}
\eeq
decreases in a more or less monotonous manner for the majority of gauge fixing 
trials until the condition (\ref{eq:gaugefixstop}) becomes satisfied with 
a unique $\epsilon_{lor}=10^{-7}$.
A monotonous OR behavior is reasonably satisfied for a final lower SA 
temperature value $~T_{\rm final}=0.01~$~\cite{Schemel:2006da}. The number of 
temperature steps has been chosen to be $1000$ for the smaller
lattice sizes and increased to $2000$ for the lattice size $30^4$ and bigger.
The finalizing OR algorithm using the standard Los Alamos type overrelaxation
with the parameter value $\omega = 1.7$ requires typically a number of
iterations varying from $O(10^2)$ to $O(10^3)$. 

In what follows we call the combined algorithm employing SA (with 
finalizing OR) and $Z(2)$ flips the `FSA' algorithm. By repeated starts of 
the FSA algorithm we search in each $Z(2)$ Polyakov loop sector several times for 
the best (``\bc'') copy. The total number of copies per configuration
$N_{copy}$ generated for each $\beta$-value and lattice size is indicated
in \Tab{tab:data_sets}. In order to demonstrate the Gribov copy effect
we can compare with the results obtained from the randomly chosen first
(``\fc'') copy.  

Some more details to speed up the
gauge fixing procedure are described in \cite{Bornyakov:2008yx}.

\section{Influence of Gribov copies on the gluon propagator}
\label{sec:gribov}

Our efficient gauge fixing procedure changes significantly the momentum
dependence of the gluon propagator $D(p)$ in the IR-region in comparison
with the result of the standard overrelaxation method. 

\begin{figure}[tb]
\centering
\includegraphics[width=7.1cm,angle=270]{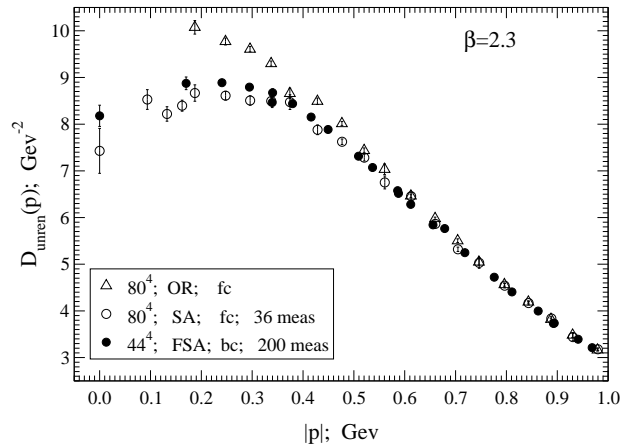}
\caption{Comparison of data obtained for  
\bc FSA gauge fixing with those obtained with the standard 
\fc OR - method and the \fc SA algorithm (all for $\beta=2.30$).
}
\label{fig:OR_vs_SA}
\end{figure}

In \Fig{fig:OR_vs_SA} we compare our \bc FSA results for the 
unrenormalized gluon propagator $D(p)$ calculated on a $44^4$ lattice 
with those of the standard \fc OR method obtained for an $80^4$  lattice
in Ref. \cite{Sternbeck:2007ug};  all data produced for $\beta=2.3$
\footnote{Only momenta $p \ageq 0.2$ GeV are shown for data from
\cite{Sternbeck:2007ug}, note also that we are using a slightly different
value for the lattice spacing than in \cite{Sternbeck:2007ug}.}. 
One can see that our \bc FSA data points lie essentially below those of
the \fc OR method for momenta $|p| \aleq 0.7$ GeV.  Thus, we observe
that the OR method with one gauge copy produces 
unreliable results for this range of momenta. Note that our lattice size 
is approximately twice as small as that in Ref. \cite{Sternbeck:2007ug}.

In the same figure we compare also with the \fc SA results  (no flips
taken into account but much longer SA schedule applied)  obtained for an
$80^4$ lattice in Ref. \cite{Bogolubsky:2009qb}. The data look consistent
for the momentum region $|p| \ageq 0.3$ GeV. At smaller momenta, e.g.,
at $|p| \aleq 0.3$ GeV the \fc SA data points show the tendency to lie at
somewhat lower values. The reason for this difference  can be attributed
to finite-volume effects (our lattice size is much smaller)  or to
uncertainties due to the lower statistics in \cite{Bogolubsky:2009qb}. 
In any case we confirm that the $Z(2)$ flips have the tendency to 
lower finite-size effects \cite{Bogolubsky:2007bw}. 

Let us define the Gribov copy sensitivity parameter $\Delta(p)$ as a
normalized difference of the \fc and \bc gluon propagators
\beq
\Delta(p) = \frac{D^{fc}(p) - D^{bc}(p)}
{D^{bc}(p)}~,
\label{delta}
\eeq
where the numerator has been obtained by averaging the differences 
between \fc SA and \bc FSA propagators calculated for every 
configuration and normalized with the \bc (averaged) propagator.

\begin{figure}[tb]
\centering
\includegraphics[width=7.1cm,angle=270]{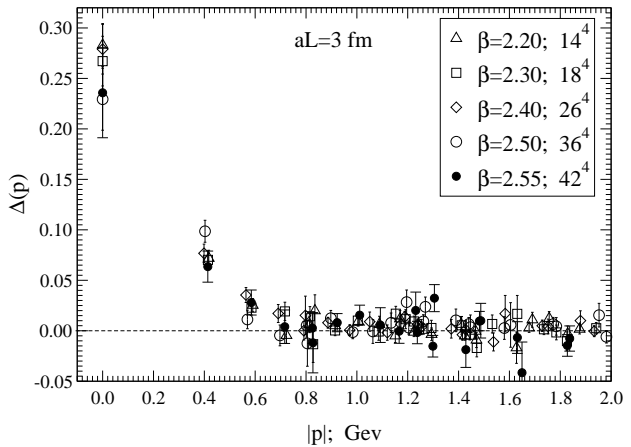}
\caption{The momentum dependence of the Gribov copy sensitivity
parameter $\Delta(p)$ for various lattices with physical size 
$aL \simeq 3$ fm.
}
\label{fig:del_v3}
\end{figure}

\begin{figure}[tb]
\centering
\includegraphics[width=7.1cm,angle=270]{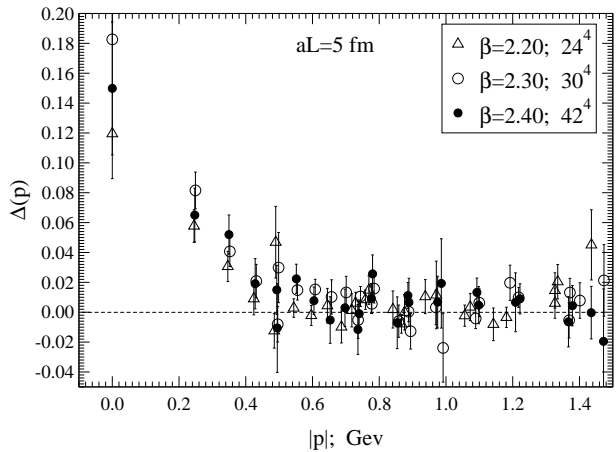}
\caption{The momentum dependence of $\Delta(p)$ for various 
lattices with physical size $aL \simeq 5$ fm.
}
\label{fig:del_v5}
\end{figure}

\begin{figure}[tb]
\centering
\includegraphics[width=7.1cm,angle=270]{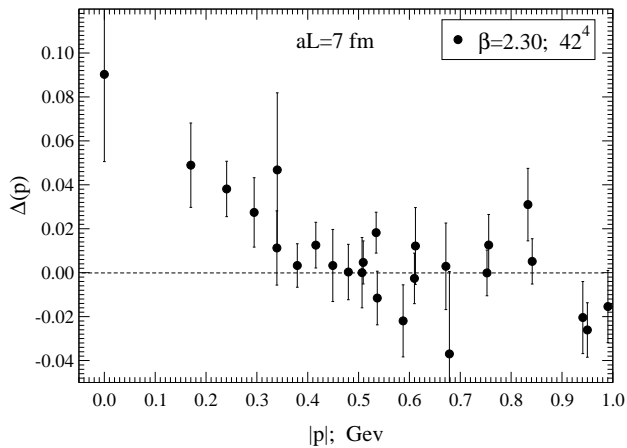}
\caption{The momentum dependence of $\Delta(p)$ on a 
lattice with physical size $aL \simeq 7.3$ fm.
}
\label{fig:del_v7}
\end{figure}

In Figs. \ref{fig:del_v3}, \ref{fig:del_v5}, and \ref{fig:del_v7} 
we show the momentum dependence of the Gribov copy sensivity 
parameter $\Delta(p)$ for different lattices with 
physical sizes $aL \simeq 3$ fm, $aL \simeq 5$ fm and $aL \simeq 7.3$
fm, respectively.

Evidently, the Gribov copy effect is rather strong in the deep
IR-region. It does not disappear with rising $\beta$, i.e. with  the
lattice spacing $a$ becoming smaller. However, the effect decreases
rapidly for rising momentum. Moreover, it is encouraging to see that the
influence of Gribov copies for fixed momentum demonstrates clear 
tendency to decrease  with increasing {\it physical} size $aL$.  Thus, 
with our gauge fixing procedure we do not find any Gribov copy effect 
for momenta $p \ageq 0.6$ GeV on lattices with $aL \simeq 3$ fm, 
while on lattices with $aL \simeq 7.3$ fm the gluon propagator for 
all momenta  $p \ageq 0.4$ Gev  is free of Gribov copy effect. 
This tendency is in accordance with a conjecture by Zwanziger 
in~\cite{Zwanziger:2003cf} and  was seen already for smaller lattice 
sizes in~\cite{Bogolubsky:2005wf}.

Let us finally note that the above observations are valid for our FSA
gauge fixing method (employed with rather large number of gauge copies),
which is proved to be much more powerful than, e.g., standard OR method.
Of course, we cannot exclude that other even more efficient gauge 
fixing methods can be invented which might bring us even closer to 
the global extremum of the gauge functional. But we are convinced
that this will not change the conclusions given here.

We would like to emphasize that the supposition to find the
Gribov copies as close as possible to the global extremum of the 
Landau gauge functional differs substantially from the recent claim
to study those Gribov copies maximally enhancing the infrared asymptotics 
of the ghost dressing function \cite{Maas:2009se}. In as far these two
different strategies really provide the different {\it decoupling} and
{\it scaling} solutions, respectively, for the gluon and ghost 
propagators within the thermodynamic limit remains an interesting 
question. 

\section{Numerical gluon propagator results}
\label{sec:results}

In order to suppress lattice artifacts from the beginning 
we followed Ref.~\cite{Leinweber:1998uu} and 
selected the allowed lattice momenta as surviving
the {\it cylinder cut} 
\beq
\sum_\mu k^2_\mu - \frac{1}{4}(\sum_\mu k_\mu )^2 \leq 3 \,.
\eeq
Moreover, we have applied the  ``$\alpha$-cut'' \cite{Nakagawa:2009zf}
$p_{\mu} \le (2/a) \alpha$ for every component, in order to
keep close to a linear behavior of the lattice momenta 
$p_{\mu}=(2 \pi k_{\mu})/(aL), ~~k_{\mu} \in (-L/2,L/2]$. 
We have chosen $\alpha=0.5$. Obviously, this cut influences large 
momenta only.

We define the renormalized propagator $D_{ren}(p)$ according to
the momentum subtraction schemes (MOM) by
\bea
D_{ren}(p,~\mu) ~&=& {\cal Z}_{ren}(\mu,1/a) ~D(p, 1/a) \\
D_{ren}(p=\mu) &=& 1 / \mu^2\,.
\label{eq:renorm}
\eea
In practice we have fitted the bare propagators $D(p, 1/a)$ 
with an appropriate function (see \Eq{eq:fitfunction} below) and then
used the fits for renormalizing $D(p)$.
But it has to be seen, that multiplicative renormalizability  really
holds in the non-perturbative regime. For this it is sufficient to
prove, that ratios of the renormalized (or  unrenormalized) propagators
obtained from different cutoff values  $1/a(\beta)$ will not depend on
$p$ at least within a certain  momentum interval $p_{min}, p_{max}$,
where $p_{max}$ should be the maximal momentum surviving all the cuts
applied.  

\begin{figure}[tb]
\centering
\includegraphics[width=7.1cm,angle=270]{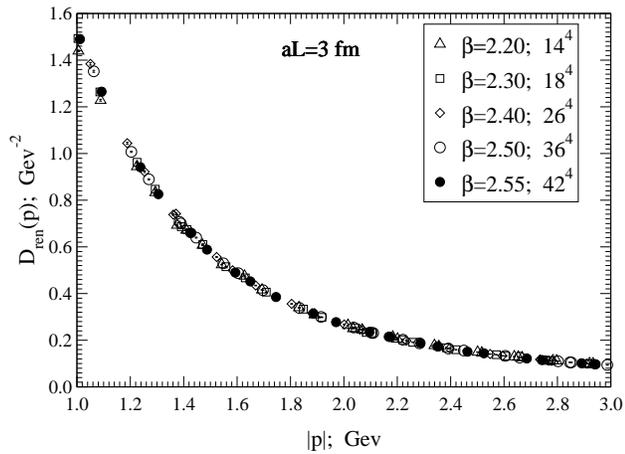}
\caption{The momentum dependence of the gluon propagator $D_{ren}(p)$ 
for five different lattice spacings and $|p| \ageq 1$ GeV.
The physical linear box size is $aL \simeq 3$ fm.
}
\label{fig:glp_ren_v3_UV}
\end{figure}

In what follows the subtraction momentum has always
been chosen to be $\mu=2.2$ GeV.  In \Fig{fig:glp_ren_v3_UV} we show
the momentum dependence of the renormalized gluon propagator $D_{ren}(p)$
at comparatively large momenta ($|p| \ageq 1$ GeV) for five different
lattice spacings but with (approximately) the same physical size $aL
\simeq 3$ fm  (for exact values see \Tab{tab:data_sets}). Evidently,
in this momentum range the finite-spacing effects are rather small,
at least for inverse bare coupling values $4/g_0^2 \equiv \beta > 2.2$.

\begin{figure}[tb]
\centering
\includegraphics[width=7.1cm,angle=270]{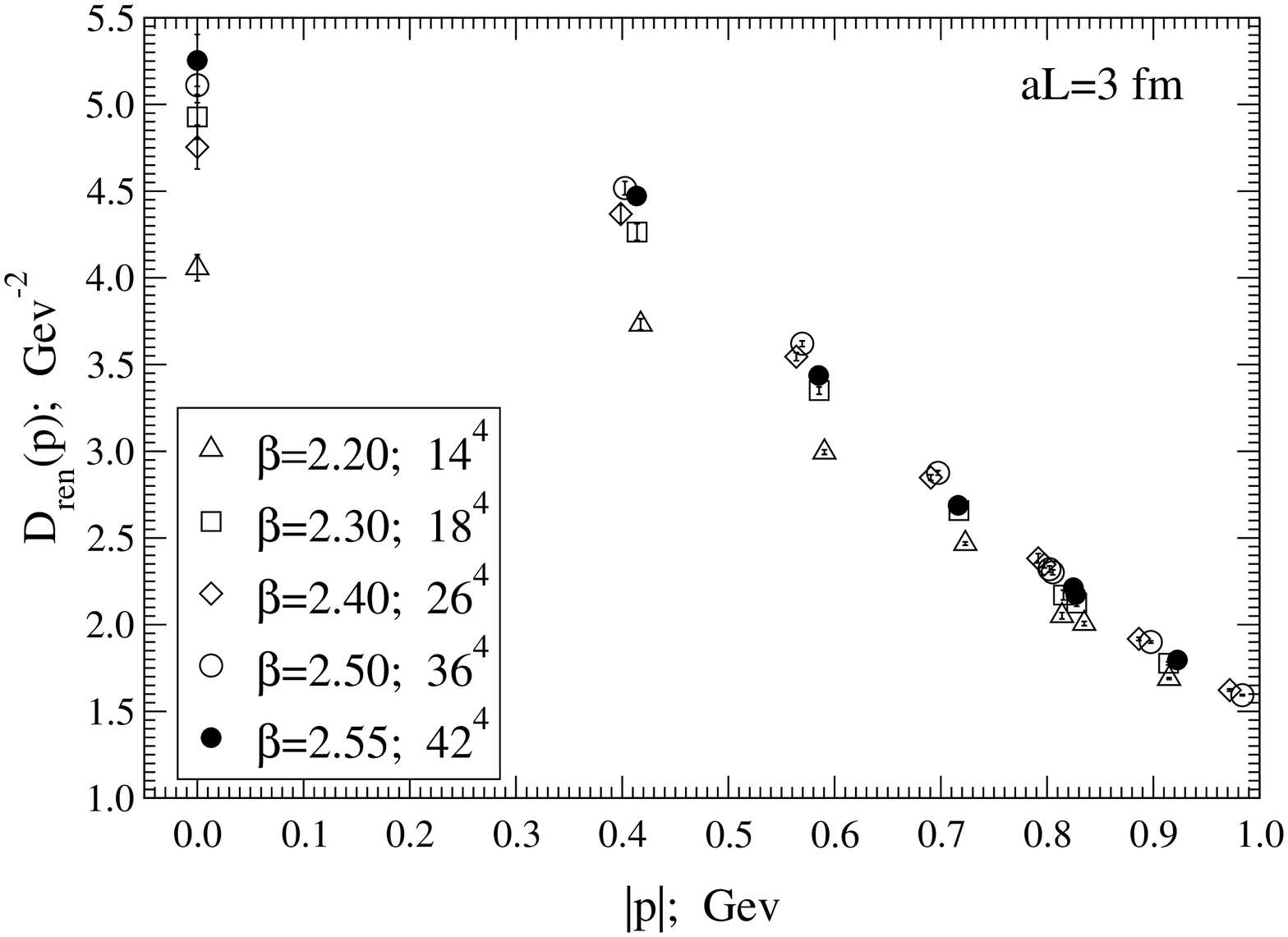}
\caption{The same as in \Fig{fig:glp_ren_v3_UV}
but for $|p| \aleq 1$ GeV.
}
\label{fig:glp_ren_v3_IR}
\end{figure}

\begin{figure}[tb]
\centering
\includegraphics[width=7.1cm,angle=270]{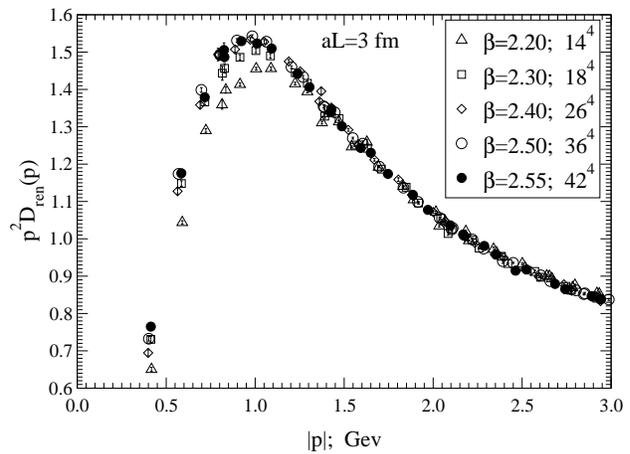}
\caption{The momentum dependence of the renormalized dressing function 
$Z_{ren}(p)$ for five different lattice spacings. 
The physical linear box size is $aL \simeq 3$ fm.
}
\label{fig:glz_ren_v3}
\end{figure}

In \Fig{fig:glp_ren_v3_IR} for the same data set, we show the
IR region only, whereas \Fig{fig:glz_ren_v3} presents the momentum dependence 
of the renormalized gluon dressing function $Z_{ren}(p)=p^2 D_{ren}(p)~$ 
for the same lattice spacings as in Figs. \ref{fig:glp_ren_v3_UV} and 
\ref{fig:glp_ren_v3_IR}. In both cases we see that there are quite strong 
deviations at least for $\beta=2.2$. The early turnover of the gluon 
propagator $D_{ren}(p)$ to an IR flattening is much less pronounced for higher 
$\beta-values$ and thus, has to be attributed to lattice artifacts. We have
to suspect that similar computations done for $SU(3)$ at the quite low
value $\beta=5.7$ suffer from the same problem~\cite{Bogolubsky:2009dc}. 

\begin{figure}[tb]
\centering
\includegraphics[width=7.1cm,angle=270]{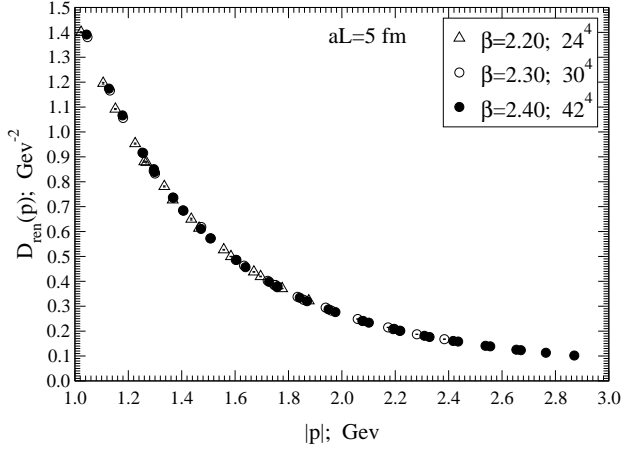}
\caption{The momentum dependence of the renormalized gluon propagator 
$D_{ren}(p)$ for three different lattice spacings and $|p| \ageq 1$ GeV. 
The physical linear box size is $aL \simeq 5$ fm.
}
\label{fig:glp_ren_v5_UV}
\end{figure}

\begin{figure}[tb]
\centering
\includegraphics[width=7.1cm,angle=270]{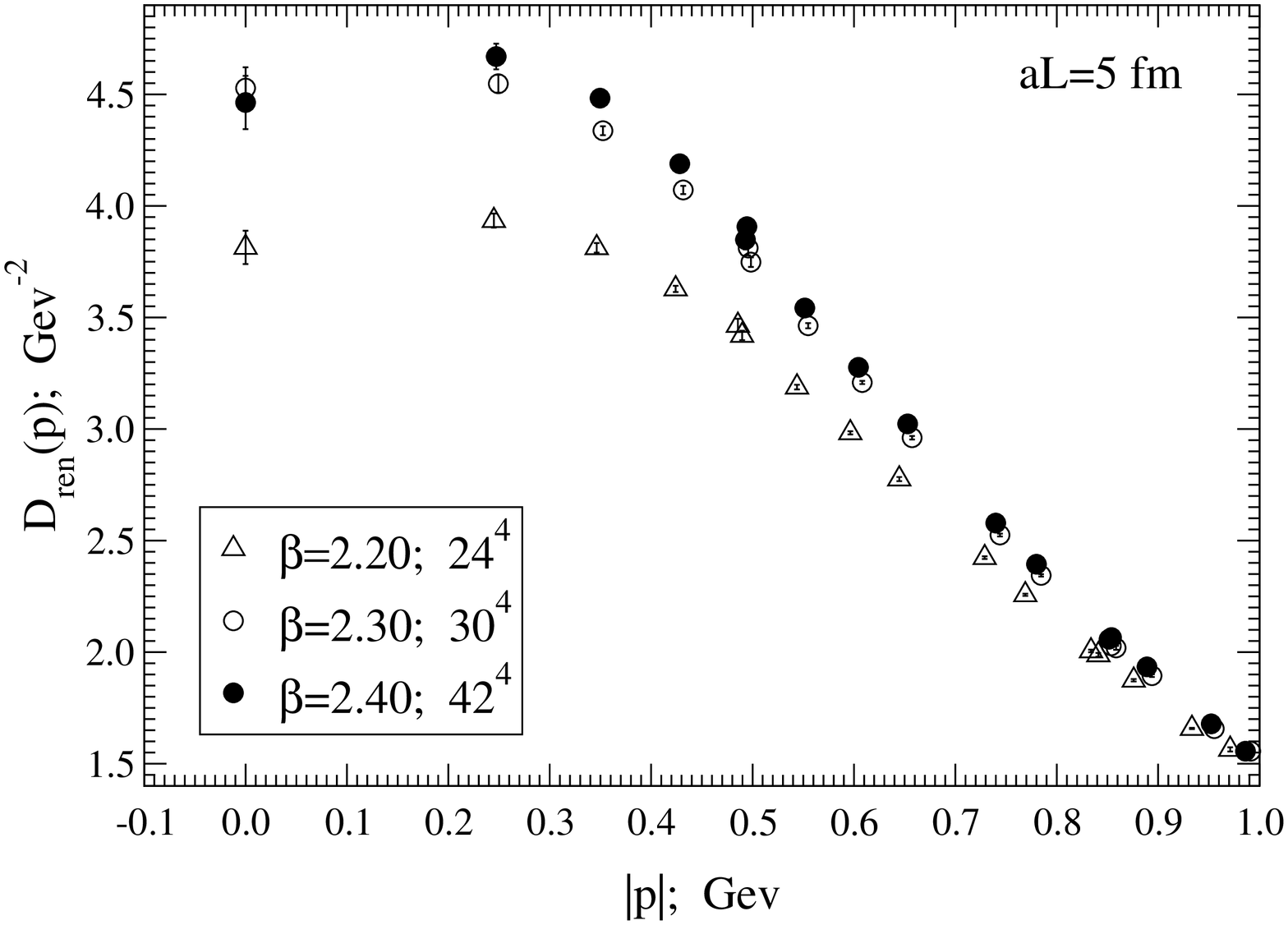}
\caption{The same as in \Fig{fig:glp_ren_v5_UV}
but for $|p| \aleq 1$ GeV.
}
\label{fig:glp_ren_v5_IR}
\end{figure}

\begin{figure}[tb]
\centering
\includegraphics[width=7.1cm,angle=270]{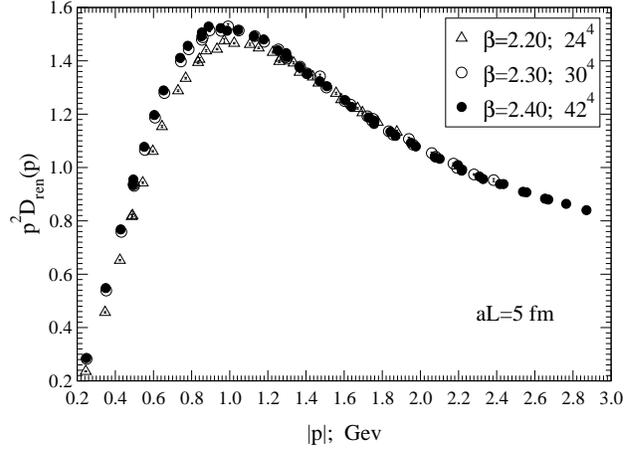}
\caption{The momentum dependence of the renormalized dressing function 
$Z_{ren}(p)$ for three different lattice spacings. 
The physical linear box size is $aL \simeq 5$ fm.
}
\label{fig:glz_ren_v5}
\end{figure}

In Figs. \ref{fig:glp_ren_v5_UV}, \ref{fig:glp_ren_v5_IR}, and 
\ref{fig:glz_ren_v5} we show an analogous set of data but now for
the larger volume of about $~(5 \mathrm{fm})^4$. Also in this case
we see that the continuum limit is fastly reached for $~\beta \geq 2.3~$
in the whole momentum range, whereas for $~\beta=2.2~$ lattice artifact  
deviations occur, which become particularly strong in the infrared 
region $|p| \aleq 1$ GeV. 


\begin{figure}[tb]
\centering
\includegraphics[width=7.1cm,angle=270]{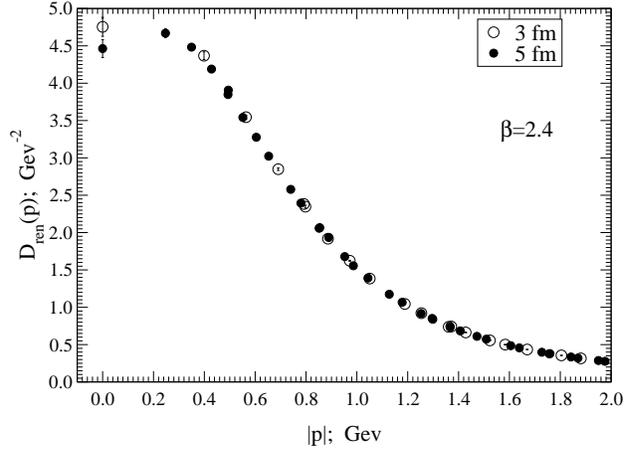}
\caption{The momentum dependence of the renormalized gluon propagator
$D_{ren}(p)$ at fixed $\beta=2.4$ for two different physical volumes.
}
\label{fig:glp_ren_v3_v5}
\end{figure}

\begin{figure}[tb]
\centering
\includegraphics[width=7.1cm,angle=270]{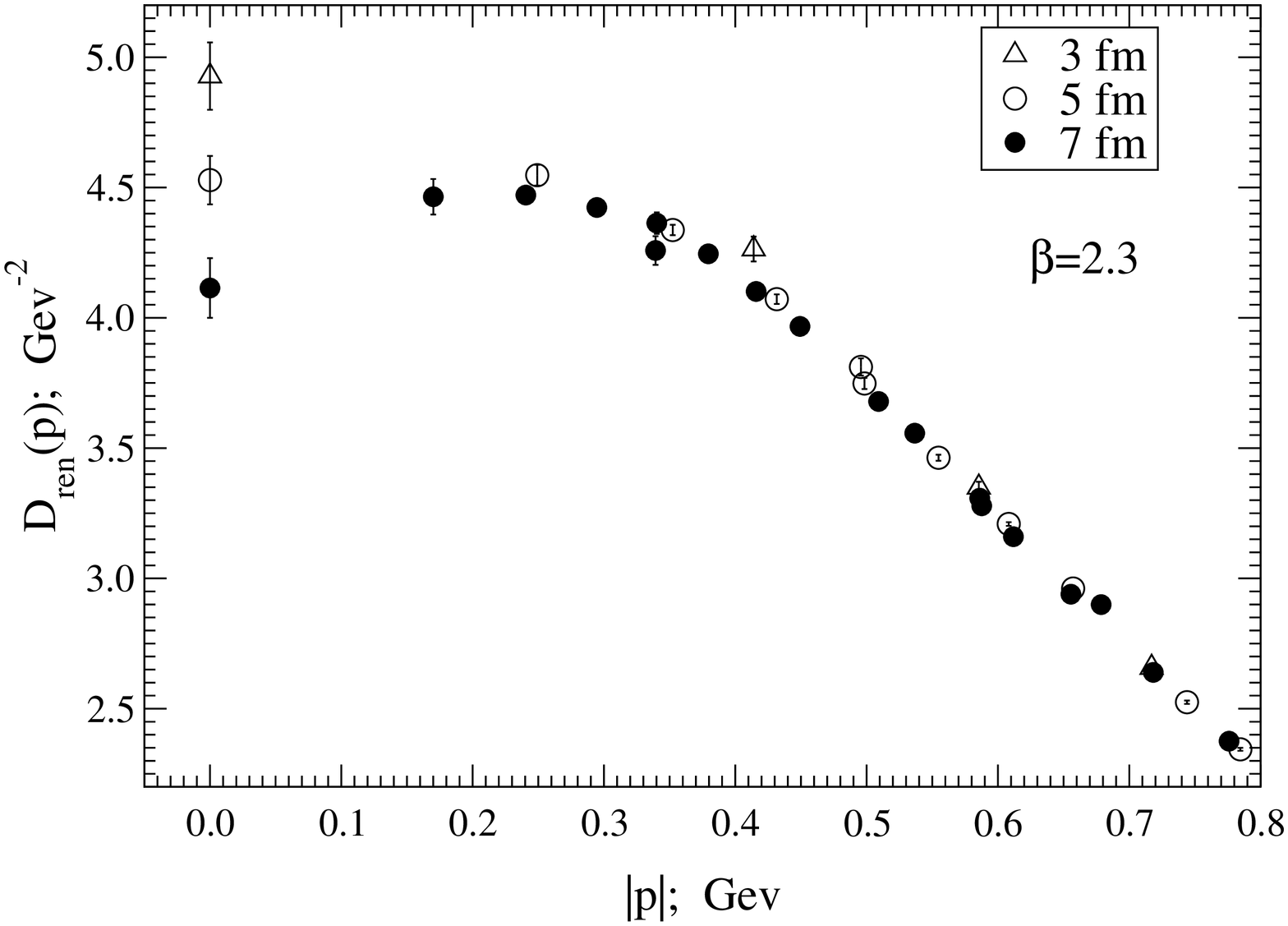}
\caption{The momentum dependence of the renormalized gluon propagators 
$D_{ren}(p)$ at fixed $\beta=2.3$ for three different physical volumes.
}
\label{fig:glp_ren_v3_v5_v7}
\end{figure}

In order to estimate finite-volume effects we compare our \bc
FSA data for the propagator $D_{ren}(p)$ obtained for the same
$\beta$-values but different $aL$. In \Fig{fig:glp_ren_v3_v5} we
show the momentum dependence of the renormalized gluon propagators
$D_{ren}(p)$ for two different physical sizes $aL$ at $\beta=2.4$,
and in \Fig{fig:glp_ren_v3_v5_v7} $D_{ren}(p)$ is presented for three
different volumes at the somewhat stronger coupling $\beta=2.3$. 
One can see that finite-volume effects are present only for the 
zero and minimal nonzero momenta and in the latter case they are 
rather small. Moreover, the IR  flattening becomes visible only 
for the largest volume $aL \simeq 7$ fm.

Evidently, our data for the zero-momentum propagator $D_{ren}(p=0)$ 
show a clear tendency to decrease with increasing physical size $aL$.
However, it remains difficult to speculate about the infinite volume
limit of $D_{ren}(0)$.

\begin{figure}[tb]
\centering
\includegraphics[width=7.1cm,angle=270]{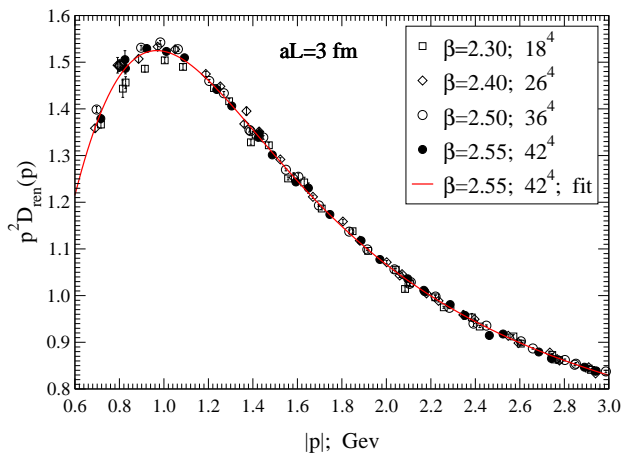}
\caption{The momentum dependence of the renormalized dressing 
function $Z_{ren}(p)$ fitted at the smallest available lattice 
spacing compared with the data of coarser lattices. 
The linear box size is $aL \simeq 3$ fm.
}
\label{fig:glz_ren_v3_fit}
\end{figure}

\begin{figure}[tb]
\centering
\includegraphics[width=7.1cm,angle=270]{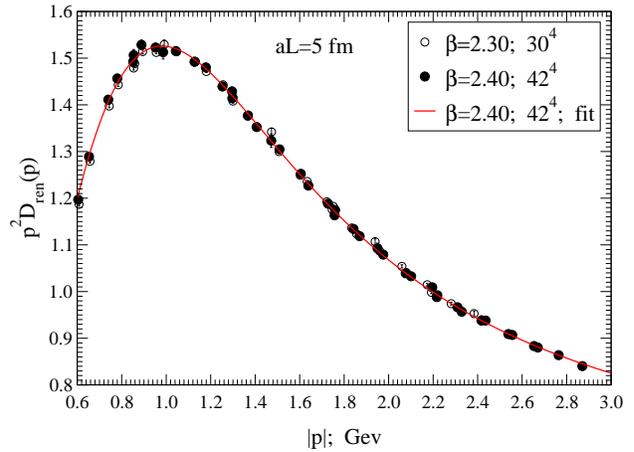}
\caption{The momentum dependence of the renormalized dressing 
function $Z_{ren}(p)$ as for \Fig{fig:glz_ren_v3_fit} but
with linear box size $aL \simeq 5$ fm.
}
\label{fig:glz_ren_v5_fit}
\end{figure}

\begin{figure}[tb]
\centering
\includegraphics[width=7.1cm,angle=270]{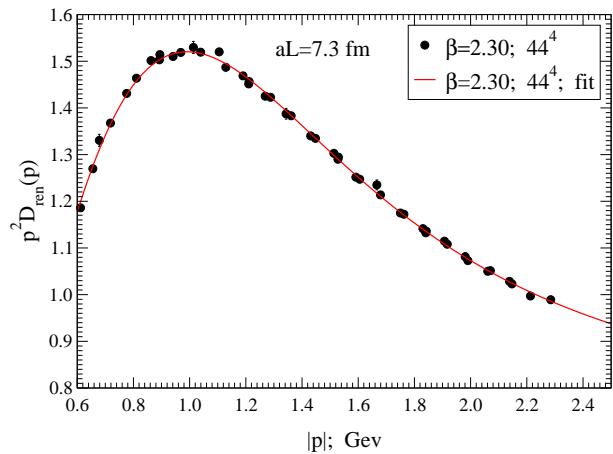}
\caption{The momentum dependence fit of the renormalized dressing 
function $Z_{ren}(p)$ for $\beta=2.3$ and linear box size 
$aL \simeq 7$ fm.
}
\label{fig:glz_ren_v7_fit}
\end{figure}

In the literature one can find quite a few functional forms suggested
to describe the gluon propagator in the IR-region, most of them of
purely phenomenological origin, see e.g. \cite{Leinweber:1998uu}. For
$\beta \geq 2.3$ we have fitted the momentum dependence of the gluon
propagator with an ansatz describing a massive behavior in the infrared
\beq
D_{ren}(p) = \frac{a_1}{p^2+m^2} + \frac{a_2}{(p^2+m^2)^2}
+ \frac{a_3}{(p^2+m^2)^4}~,
\label{eq:fitfunction}
\eeq
\noi $a_1,a_2,a_3$ and $m$ are fit parameters.
The fitting curves allow to compare quite easily the propagators 
obtained on different volumes.  We show the fit results 
obtained for $ p \ageq 0.6$ GeV
in Figs. \ref{fig:glz_ren_v3_fit}, \ref{fig:glz_ren_v5_fit}, and 
\ref{fig:glz_ren_v7_fit} for the same three volumes as considered 
before. The fit parameters are provided in \Tab{tab:fit}. 

\begin{figure}[tb]
\centering
\includegraphics[width=7.1cm,angle=270]{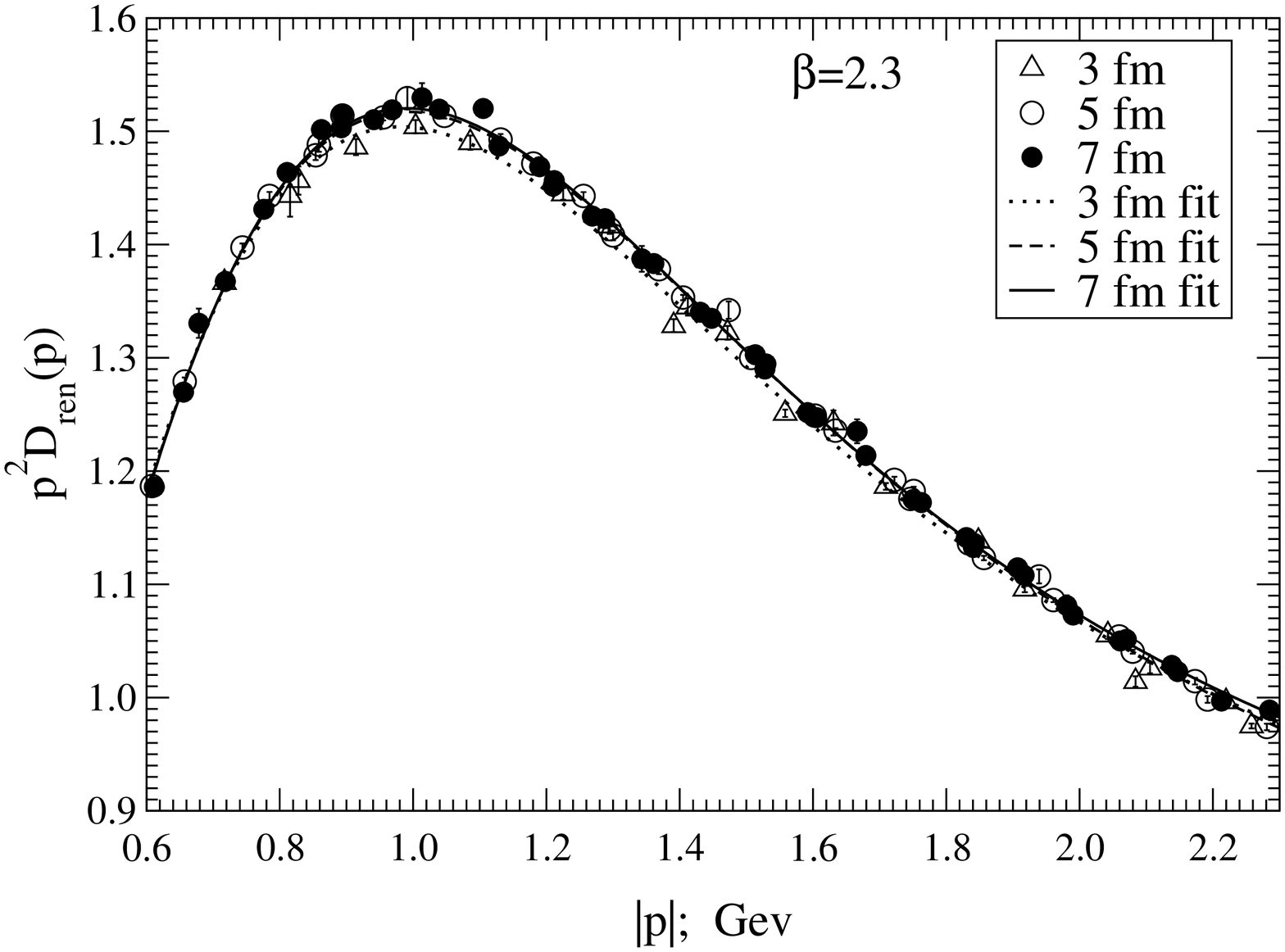}
\caption{The renormalized dressing functions 
$Z_{ren}(p)$ at $\beta=2.3$. The data and the corresponding fits 
are shown for linear box sizes $aL \simeq 3,~5,~7$ fm.
}
\label{fig:glz_ren_v3_v5_v7_fit}
\end{figure}

\begin{figure}[tb]
\centering
\includegraphics[width=7.1cm,angle=270]{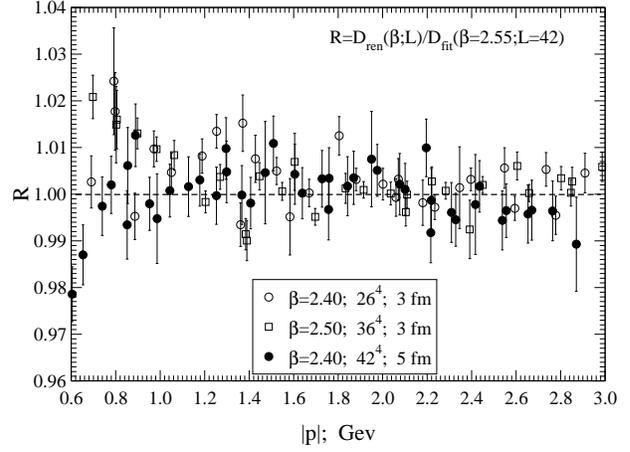}
\caption{The momentum dependence of the $R$-ratio for different 
$\beta$ values and physical volumes.
}
\label{fig:ratio_fit}
\end{figure}

In \Fig{fig:glz_ren_v3_v5_v7_fit} we compare the data together 
with the corresponding fit curves obtained at $~\beta=2.3~$ 
for the three physical volumes. We see that in the range shown
$~p~\geq~0.6~\mathrm{GeV}~$ finite-size effects are small at least
for physical linear lattice sizes $~aL~\geq~5$ fm.
%
\begin{table}[h]
\begin{center}
\vspace*{0.2cm}
\begin{tabular}{|c|c|c|c|c|c|c|} \hline
$\beta$ & $L$ & $a_1$ & $a_2$ & $a_3$ & $m$ & $\chi^2_{df}$ \\ 
\hline\hline
2.20  & 14 & 0.62 & 4.58 &  49.5 & 1.38 &  5.0 \\
2.30  & 18 & 0.61 & 4.03 &  54.5 & 1.37 &  3.8 \\
2.40  & 26 & 0.56 & 4.31 &  50.8 & 1.35 &  1.5 \\
2.50  & 36 & 0.55 & 4.30 &  39.5 & 1.30 &  1.5 \\
2.55  & 42 & 0.56 & 4.34 &  45.0 & 1.34 &  1.2 \\
\hline\hline
2.20  & 24 & 1.02 & 1.33 & 128.4 & 1.53 &  3.6 \\
2.30  & 30 & 0.59 & 4.19 &  61.0 & 1.39 &  2.6 \\
2.40  & 42 & 0.55 & 4.39 &  53.9 & 1.37 &  0.5 \\
\hline\hline
2.30  & 44 & 0.67 & 3.65 &  72.2 & 1.41 &  1.6 \\
\hline\hline
\end{tabular}
\end{center}
\caption{Values of the fit parameters in physical units 
(with dimension $[a_1]=\mathrm{GeV}^0, 
[a_2]=\mathrm{GeV}^2, [a_3]=\mathrm{GeV}^6,
[m]=\mathrm{GeV}$) and the corresponding $\chi^2_{df}$.
} 
\label{tab:fit}
\end{table}
Finally we check the multiplicative renormalizability by 
presenting the data for the gluon propagator ratio 
\beq
R=D_{ren}(p;\beta;L)/D_{ren}^{fit}(p;\beta=2.55;L=42) 
\label{eq:R_ratio}
\eeq
in \Fig{fig:ratio_fit}. The relative deviations are below 3 percent
such that we can say multiplicative renormalizability is safe for 
$\beta \geq 2.4$ and for the momentum range $p \geq 0.6$ GeV.

\vspace{2mm}

\section{Deep infrared region and infrared mass scale}
\label{sec:infrared}

In the previous section we discussed our fits of the propagator in the
range $ 0.6  \,\aleq p \,\aleq \, 3$ GeV. An even more phenomenologically
interesting range is the deep infrared region $|p| \, \aleq \, 0.6$
GeV. As it was discussed in \Sec{sec:introduction} the infrared mass scale
is an important parameter for phenomenological analyses. Lattice
results in principle can provide model independent information on this
subject. Our results obtained in this region for $aL = 5$ fm and $\beta=2.3$
and $2.4$ exhibit only a weak dependence on the lattice spacing
(see Figs. \ref{fig:glp_ren_v5_IR}, \ref{fig:glz_ren_v5}). Moreover, the finite-size
dependence of $D(0)$ seems to be moderate at $\beta=2.4$ (compare with
Figs. \ref{fig:glp_ren_v3_v5}, \ref{fig:glp_ren_v3_v5_v7}). Therefore, let
us speculate that our results for momenta below $0.6$ GeV for $\beta=2.4$ 
and $aL = 5$ fm are already close to the continuum and to the large-volume 
limit and that multiplicative renormalization can be assumed, too. 
Under these assumptions we have fitted our results at $p\, \aleq \, 0.6$
GeV separately. We used two fitting functions to fit our data in this
range of momenta. The first one is just the pole-type propagator providing 
an effective gluon mass scale $m_g$ for $p \to 0$

\beq
D_\mathrm{pole}(p) = \frac{A}{p^2+m^2_g}\,.
\label{eq:freescalar}
\eeq

\noi The other one is of the form of a Gaussian

\beq
D_\mathrm{gauss}(p) = B e^{-(p-p_0)^2/m^2_\mathrm{IR}}\,.
\label{eq:gauss}
\eeq

\begin{figure}[tb]
\centering
\includegraphics[width=7.1cm,angle=270]{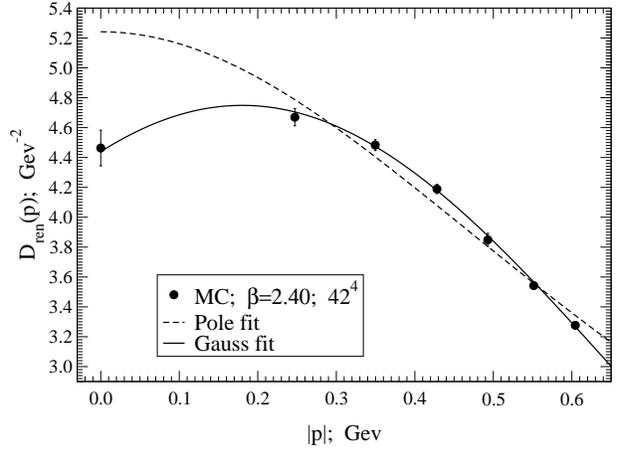}
\caption{The renormalized gluon propagator at $\beta=2.4$ on $aL=5$ fm lattice 
together with the infrared fits acc. to Eqs. 
(\ref{eq:freescalar},\ref{eq:gauss}).
}
\label{fig:infrared_fit}
\end{figure}

\noi $A, m_g, B, p_0, m_\mathrm{IR}$  are fit parameters. The
results obtained for $\beta=2.4$ and $aL=5$ fm are compared in 
\Fig{fig:infrared_fit}. 

One can see that the pole-type momentum dependence given in
\Eq{eq:freescalar} is not suitable for fitting the gluon
propagator in the considered range of momenta. This is supported by
the large value of the $\chi/ndf$ parameter value which was found to
be about 14 if $p=0$ is included and about 7 if this point is
excluded. Therefore, the definition of an effective gluon mass $m_g$
via the tree-level \Eq{eq:freescalar} turns out to be problematic. 

On the contrary, the Gaussian-type momentum dependence given in
\Eq{eq:gauss} nicely fits the data for the gluon propagator in
the range  $p \, \aleq \, 0.6$ GeV  with $\chi^2/\mathrm{ndf}$ about
0.7 (0.5) with $p=0$ included (excluded). Therefore, the Gaussian
functional form allows to define an infrared (mass) scale 
$m_\mathrm{IR}$. The numerical fit result is $m_\mathrm{IR}=0.69(3)$ 
MeV for $p=0$ taken into account in the fit and 0.68(4) MeV, if 
$p=0$ is ignored. 

It is also interesting to note that our data suggest the existence
of a maximum of the gluon propagator at a non-zero momentum
$p_0$. We have already observed this for $SU(2)$ in four dimensions  
in our previous work \cite{Bornyakov:2008yx}, where we studied the 
propagator on large lattices at $\beta=2.2$, i.e. rather far 
from the scaling region. Our new results reported for $\beta=2.3$ 
and 2.4 (see Figs.~\ref{fig:glp_ren_v3_v5},\ref{fig:glp_ren_v3_v5_v7})
show that the maximum seems to persist in the continuum limit. 
With our fit function \Eq{eq:gauss} we obtained its position at
$p_0=180(15)$ MeV. The mere existence of the maximum already  
contradicts a simple effective gluon mass prescription 
$\sim 1/(p^2 + m_g^2$.
 
\section{Conclusions}
\label{sec:conclusions}

In this work we investigated numerically the renormalized Landau
gauge gluon propagator $D_{ren}(p)$ in the pure gauge  $SU(2)$ lattice
theory. The main goal of this study was to study the approach to the
continuum limit, especially in the infrared region $|p| \aleq 1$ GeV.

In order to disentangle finite-spacing from finite-volume effects 
we calculated the propagators on lattices with {\it physical} size  
$aL$ equal approximately 3 fm at various $\beta$-values in the range
of $\beta=2.2, \ldots, 2.55$ and on lattices with $aL \approx 5$ fm
for $\beta=2.2, \ldots, 2.4$ (see \Tab{tab:data_sets} for 
details).  Calculations were made also on $aL \approx 7.3 $fm lattices 
at $\beta=2.3$.  Our lattice volumes  varied from $L^4=14^4$ to $44^4$.
For physical volumes $(aL)^4 \simeq (3 \mathrm{fm})^4$ and 
$(5 \mathrm{fm})^4$, we have checked
the scaling behaviour assuming the lattice spacing $a$ to depend on
$\beta$ as determined from the string tension. The comparison of the
renormalized propagators calculated for different physical volumes
then allowed to estimate the influence of the finite (physical) volume
in the infrared regime.

Special attention has been paid to the dependence on the choice of 
Gribov copies. In our previous papers  
\cite{Bogolubsky:2007bw,Bogolubsky:2008mh} we have seen that the 
finite-volume behavior of the gluon propagator (and not only
of the ghost propagator) is sensitive to the way how the Landau gauge
is fixed. We found indications that by enlarging the gauge orbits by 
$Z(2)$-flip operations and by applying the simulated annealing method 
with consecutive overrelaxation (`FSA' algorithm) the volume 
dependence becomes suppressed.
However, our former investigations went immediately to largest accessible
lattice volumes by employing coarse lattices ($\beta=2.2$), such 
that the continuum limit remained to be studied. 

\vspace{1mm}

Our findings can be summarized as follows.

\vspace{1mm}

\noi {\bf 1)} In the region $|p| \ageq 0.6$ GeV we observe very nice agreement
between renormalized propagators $D_{ren}(p)$ obtained for different
lattices with $\beta \geq 2.4$. For larger momenta scaling holds for even 
smaller $\beta$ values. 
In contrast, in the deep infrared region ($|p| \aleq 0.6$ GeV) the 
scaling violation is quite strong, especially for $\beta=2.2$.  However, 
with increasing $\beta$ finite-spacing effects rapidly decrease: our data
seem to `converge' to some limiting curve of the renormalized propagator
$D_{ren}(p)$ (compare, e.g., propagators at $\beta=2.2$ and $\beta=2.55$),
thus indicating the approach to the {\it continuum limit} in the
fixed physical volume $(aL)^4$.

\vspace{1mm}

\noi {\bf 2)} 
Using our gauge fixing procedure we observed finite (physical)
volume effects only for zero and minimal nonzero momenta. We made
this observation for $\beta=2.2$ in \cite{Bornyakov:2008yx}, here we
confirm it for $\beta=2.3$ and 2.4 and thus it can be extended to the
continuum limit.

\vspace{1mm}

\noi {\bf 3)}
In our previous papers \cite{Bogolubsky:2007bw,Bornyakov:2008yx}
we calculated gluon propagators on various lattices at $\beta=2.2$.
We observed the appearance of a maximum at a non-zero value of the
momentum $p$ on lattices with comparatively large volume ($aL \ge 6.7$
fm). Our new data obtained at larger $\beta$-values do confirm this
observation. Fitting the propagator calculated at $\beta=2.4$  by the
fitting function \Eq{eq:gauss} we obtained our best estimation for
the  position of this maximum as $p_0=180(15)$ MeV. This number might
slightly change in the continuum limit. Also we observed that the
zero-momentum gluon propagator $D(0)$ has a tendency to decrease with
growing lattice size $L$. We did not try here to extrapolate its 
infinite-volume value.

\vspace{1mm}

\noi {\bf 4)}
The effective gluon mass $m_g$ has been employed as an important parameter 
in various phenomenological analyses. We fitted our results at $p
\aleq 0.6$ GeV using our data for the gluon propagator at $\beta=2.4$
on lattices with $aL \approx 5$. It was found that the pole-type
momentum dependence given by \Eq{eq:freescalar} does not provide 
an adequate description of our data at small momenta, 
which makes it problematic to define this parameter from the Landau
gauge gluon propagator.
On the contrary, a Gaussian-type behavior given by \Eq{eq:gauss} fits
the data nicely. It allows to define an alternative infrared (mass) 
scale $m_\mathrm{IR} = 0.69(3)$ GeV describing the approach to the 
infrared limit. Its consequences for phenomenological applications 
remain to be seen. 

\vspace{1mm}

\noi {\bf 5)} We confirm that the Gribov copy influence is very strong in
the deep infrared region. Comparing our \bc FSA results calculated on a
$44^4$ lattice with those of the standard \fc OR method obtained for an
$80^4$ lattice in Ref. \cite{Sternbeck:2007ug} (see \Fig{fig:OR_vs_SA})
we found that the OR method with one gauge copy produces unreliable 
results for momenta $|p| \aleq 0.7$ GeV. We conclude that
\fc OR method should be applied exclusively to large-momentum studies.
 
Our FSA method provides systematically higher values of the
gauge fixing functional as compared to the standard OR procedure.
We studied the Gribov copy effect for this method as well,
generating up to 80 gauge copies for every configuration.
We found for fixed physical volume the Gribov copy sensivity
parameter $\Delta(p)$ only weakly to depend on the lattice spacing 
$a$.  Therefore, the {\it quality} of the gauge fixing procedure in 
the study of gauge dependent observables remains important, at least, 
in the deep infrared.  

At the same time the influence of Gribov copies demonstrates
clear tendency to decrease  for fixed momentum with increasing {\it
physical} size $aL$. This tendency is in accordance with a conjecture
by Zwanziger in~\cite{Zwanziger:2003cf} and was seen already for smaller
lattice sizes in~\cite{Bogolubsky:2005wf}.

\vspace{1mm}

\subsection*{Acknowledgments}
This investigation has been partly supported by the Heisenberg-Landau
program of collaboration between the Bogoliubov Laboratory of Theoretical 
Physics of the Joint Institute for Nuclear Research Dubna (Russia) and 
German institutes and partly by the joint DFG-RFBR grant 436 RUS 113/866/0-1
and the RFBR-DFG grant 06-02-04014. VB and VM are supported by the 
grant for scientific schools NSh-679.2008.2. VB is supported by grants  
RFBR 07-02-00237-a, RFBR 08-02-00661-a and RFBR 09-02-00338-a. 


\end{document}